\documentclass[
 aip,
 amsmath,amssymb,
 reprint,
]{revtex4-2}

\usepackage{graphicx}
\usepackage{dcolumn}
\usepackage{bm}
\usepackage{siunitx}
\usepackage{hyperref}

\usepackage[utf8]{inputenc}
\usepackage[T1]{fontenc}
\usepackage{mathptmx}
\usepackage{etoolbox}

\usepackage{float}
\usepackage{xcolor}

\makeatletter
\def\@email#1#2{%
 \endgroup
 \patchcmd{\titleblock@produce}
  {\frontmatter@RRAPformat}
  {\frontmatter@RRAPformat{\produce@RRAP{*#1\href{mailto:#2}{#2}}}\frontmatter@RRAPformat}
  {}{}
}
\makeatother
\begin{document}

\preprint{AIP/123-QED}

\title{Removal of high-voltage-induced surface charges by ultraviolet light}

\author{M. T. Ziemba*}
 \altaffiliation{Present address: Imperial College London, South Kensington Campus, London SW7 2AZ, UK}
\author{J. Phrompao*}
\altaffiliation{Author to whom correspondence should be addressed: boom.phrompao@mpq.mpg.de}
\author{F. Jung}
\author{I. M. Rabey}
 \altaffiliation{Present address: Imperial College London, South Kensington Campus, London SW7 2AZ, UK}
\author{G. Rempe}

 \affiliation{Max-Planck-Institut f\"ur Quantenoptik, Hans-Kopfermann-Strasse 1, 85748 Garching, Germany}

\date{\today}

\begin{abstract}
A plethora of studies ranging from precision physics to quantum information employ ions, highly excited neutral atoms or polar molecules as tools. Motivated by the need for miniaturization, scalability and controllability, the particles are often trapped close to electrodes imprinted on dielectric surfaces. However, such geometry makes the particles susceptible to undesired surface charges. To understand how high voltage applied to the electrodes mediates the buildup of charges, we use a scanning Kelvin probe in an ultrahigh vacuum to investigate the electric potential landscape of an electrostatic molecule trap with a microstructured electrode array. We find that the surface becomes positively charged after applying voltages to the electrodes, leading to offset and patch electric potentials across the surface. Removal of these charges is demonstrated using ultraviolet light. The dynamics of this process, including the dependence on light intensity, wavelength and sample type, are investigated in detail. We find the charge removal to be faster for shorter ultraviolet wavelengths, with an additional polymer coating providing further enhancement. Our findings may be useful in a wide spectrum of experiments with electric-field-sensitive particles. \newline
\\
\noindent \textbf{The following article has been submitted to Review of Scientific Instruments.}\\
\noindent Copyright 2025 M. T. Ziemba, J. Phrompao, F. Jung, I. M. Rabey and G. Rempe. This article is distributed under a Creative Commons Attribution-NonCommercial-ShareAlike 4.0 International (CC BY-NC-SA) License.
\end{abstract}

\maketitle
\section{\label{sec:Intro}INTRODUCTION}
\noindent Trapped electric-field-sensitive particles are used in many areas of physics for basic research and technological applications. Examples include, but are not limited to, ions for quantum computing,\cite{Bruzewicz2019} Rydberg atoms for quantum simulation,\cite{Bro.20} and molecules for cold chemistry.\cite{Softley.23} As experiments advance, scalability and precise control become increasingly important. To meet the sophisticated requirements of modern experiments, more and more setups employ microfabrication techniques that allow for miniaturization of devices such as ion traps\cite{Romaszko2020} and chip traps for Rydberg atoms.\cite{Keil2016} More specifically, our own research with cold polar molecules utilizes an electrostatic trap largely consisting of two opposing glass plates on which microstructured high-voltage electrodes are deposited.\cite{Eng.11} This design allows us to trap a large number of cold polar polyatomic molecules for up to a minute. Even though the use of these microfabricated devices is indispensable for performance, it comes at the cost of an increased susceptibility of the cold molecule's electric dipole to uncontrolled electric fields originating, e.g., from charges on the insulating parts of the trap surface. \newline
\indent The problem arises when high voltage applied to surface electrodes produces charges that in turn create unknown electric potential offsets across parts of, or even the entire trap structure. These charges and corresponding potentials can build up over time and are hard to track. Spectroscopic data obtained in our experiments suggest the presence of such effects, with serious consequences such as reduced molecule cooling and inefficient extraction of ultracold molecules from the trap.\cite{Pre.16} Evidence is also obtained from experiments with ions in chip traps where performance is limited by electric-field noise at the trap frequency,\cite{Win.98} potentially related to moving patch potentials that can lead to increased motional heating of the ion in these devices.\cite{Turchette2000,Daniilidis2011} Rydberg atoms on chips can suffer from spatially inhomogeneous electric fields, too, thereby affecting their lifetime and coherence.\cite{Tau.10,Car.13} Beyond trapped particles, surface-charge buildup can also create problems in gravitational wave detectors\cite{Weiss1994,Rowan1997} and precision measurements of the Casimir force.\cite{Gar.20}
\newline
\indent Thus, significant efforts are undertaken to mitigate the adverse effects of surface charges. One strategy is to etch into the substrate between electrodes to inhibit accumulation of adsorbates.\cite{Tao.20} However, this approach does not completely eliminate surface charges and requires additional compensation voltages to be applied.\cite{Bro.07} As an alternative, several cleaning methods have been shown to successfully remove surface charges. Heat can be applied to a substrate, by baking the entire setup,\cite{Pre.18} or locally using a high-power infrared laser.\cite{Obr.07} This method can be time-consuming and may not always be practical, depending on the experiment. Argon-ion beams can also be employed to clean the device,\cite{Hit.12} but require complex equipment for creating and handling the ion beams in an ultrahigh vacuum. One simple and promising cleaning technique utilized in numerous experiments is illuminating the target with ultraviolet (UV) light. For example, heating rates of trapped ions can be reduced by cleaning the ion trap using a pulsed UV laser,\cite{All.11} the spectral shift of Rydberg atoms has been minimized by illuminating the setup with a UV diode,\cite{Pal.24} and charges on mirrors for gravitational wave detectors have been eliminated using UV light.\cite{Hew.07,Ugo.08}\newline 
\indent Inspired by these successful results, we conduct a systematic investigation into the removal of surface charge from our microstructured electrostatic trap using UV light. Towards this end, we use a scanning Kelvin probe system in an ultrahigh vacuum allowing for detailed and systematic studies of the electric-potential landscape created by high-voltage-induced charges on substrates with microfabricated electrodes. For less material usage and rapid testing, we use simplified structures as a proxy for our microstructured molecule trap.\cite{Eng.11} These comprise microstructured chromium electrodes imprinted onto a glass substrate and an identical sample coated with an additional high-voltage-resistant polymer layer. In either case, the signals from the Kelvin probe indicate the buildup of positive surface charges resulting from a high-voltage application, leading to patch potentials and offsets of the electric potential across the samples. We demonstrate successful removal of these charges by ultraviolet irradiation. Here, the effect of the irradiation by UV light of wavelengths ranging from \SI{260}{\nano\m} to \SI{410}{\nano\m} on the offset and patch electric potentials are presented. We find that faster charge removal can be achieved using UV light of a shorter wavelength for both samples. Moreover, the dependence of the dynamics of charge removal on the UV light intensity are tested for the polymer-coated sample at two wavelengths, i.e., \SI{260}{\nano\m} and \SI{370}{\nano\m}. With \SI{260}{\nano\m} and intensity higher than \SI{1.6}{\milli\watt\per\centi\m^2}, the fastest surface charge removal can be accomplished within about 20 minutes. Our findings can immediately be transferred to other systems and are therefore of broad relevance.

\section{\label{sec:Principle}PRINCIPLES OF INSULATOR SURFACE CHARGING}
\begin{figure} [h]
\centering
\includegraphics[width=0.9\hsize]{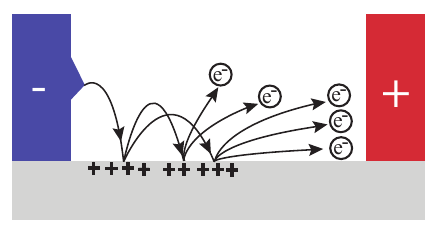}
\caption{\label{fig:Fig1_SEEA} Sketch of the charging process of an insulator-bridged electrode gap due to secondary electron emission avalanche.\cite{Boe.63,Tou.73}  The drawing shows the onset of an electron avalanche. An electron is emitted at a point due to local-electric-field enhancement created by, e.g., imperfect fabrication. The trajectories of electrons emitted from the cathode (blue) towards the insulator surface (gray) are indicated by black arrow lines. Two secondary electrons are emitted leaving a positive charge behind. These are attracted to already existing surface charges further along the insulator gap. Once having hit the insulator surface, they free another generation of secondary electrons and leave a more strongly positively charged insulator surface behind.} 
\end{figure}

\noindent The effect of insulator surface charging is a well-known phenomenon for electrodes bridged by an insulator, a geometry often encountered in surface-based microfabricated devices.\cite{Mar.11} Our microstructured molecule trap is an example of the devices with two opposite-polarity electrodes joined by an insulating surface, sketched in Fig.~\ref{fig:Fig1_SEEA}. During the operation of the trap, offset and patch electric potentials can be observed.\cite{Pre.16} These changes can be understood by considering the widely accepted model of "secondary electron emission avalanche",\cite{Boe.63,Tou.73} which suggests surface charge buildup is responsible for the observed effects. Herein, even at modest ``gross'' electric fields, e.g., locally enhanced fields suffice to extract electrons from the cathode due to field emission. While being accelerated towards the anode, a fraction of these electrons will hit the surface of the insulator. The impact leads to the emission of secondary electrons which are in turn drawn toward the anode. At this point, a positive feedback mechanism sets in. The ``seed'' electrons will leave a positive charge on the surface of the insulator, which attracts electrons being accelerated towards the anode --- be it secondary or emitted electrons at the cathode. In consequence, the number of electrons hitting the surface increases, as does the magnitude of the positive surface charge. Since the secondary electron yield is a function of the primary electron energy, an equilibrium surface-charge density will be reached, corresponding to the average electron energies at unity secondary-electron-emission yield. At this point, an electron impact will create one electron on average and the avalanche ebbs, leaving the surface-charge layer behind. Consequently, higher applied gross fields across any insulator gap should result in larger surface charge densities, because the electrons will gain on average more energy before re-attracting towards the surface. Furthermore, this type of surface charge buildup is associated with surface electric breakdown of such geometries.\cite{And.80,Neu.00,Mil.15} Thus, reducing the surface charges is crucial for assuring both the homogeneity of the electric potential and preventing damage to the device.

\section{\label{sec:Setup}EXPERIMENTAL SETUP AND SAMPLES}
\begin{figure}[h]
\centering
\includegraphics[width=1\hsize]{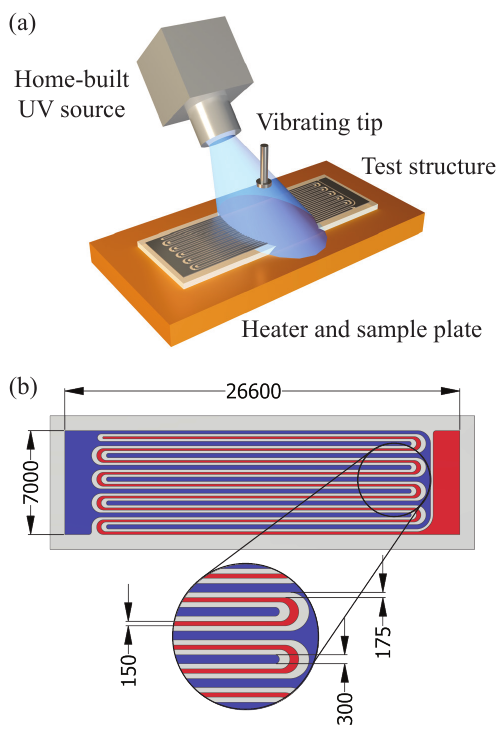}
\caption{\label{fig:Fig2_expSetup}Experimental setup and a sample. (a) Sketch of the sample on its plate underneath the tip of the Kelvin probe. A test structure is placed on a copper plate with an integrated heater at its bottom. Copper clamps to hold the sample and apply voltages are isolated from the plate by using vespel pieces (insulator). Both the copper clamps and vespel pieces are omitted in the figure for clarity. The sample is scanned by horizontally moving the sample plate underneath the Kelvin probe while the tip oscillates up and down to measure the electric potential. Ultraviolet light is introduced to the chamber from above. The drawing is not to scale as the plate is usually much larger than the sample and the ultraviolet light covers most of the test structure. (b) Drawing of the uncoated sample. The chromium electrodes are imprinted onto the AF32 glass substrate, represented as a gray area. The red and blue areas represent positive and negative electrodes, respectively. The Cyclotene$^{\text{TM}}$-coated sample is identical to the uncoated sample, but with an additional Cyclotene$^{\text{TM}}$ layer covering the whole microstructured area. Dimensions are in units of micrometers.}
\end{figure}

\noindent To observe the effects of voltage application and UV irradiation on the electric potential above a sample, we use an Ultra High Vacuum Scanning Kelvin probe system (model UHVSKP5050100) by \textit{KP Technology}. This device is sensitive to the electrochemical potential difference between a sample surface and the probe tip, which we label $\mathrm{V_{Kp}}$. By periodically changing the tip-sample spacing --- and therefore the capacitance of the system --- a measurable oscillating current flows on and off the tip. In addition to the oscillating current, a backing voltage is also applied between the sample and the tip to determine the $\mathrm{V_{Kp}}$ by using the "off-null" method.\cite{Bai.91} The $\mathrm{V_{Kp}}$ is defined negative if the sample is at a positive potential compared to the tip. It can be altered by having a voltage applied to an electrode while measuring or by having surface charges on insulator surfaces. The absolute $\mathrm{V_{Kp}}$ increases, relative to a sample with no voltages applied and no surface charges present, with either applied voltage of higher absolute value or a larger amount of surface charges of the same sign. As during measuring no voltages are applied, changes in the $\mathrm{V_{Kp}}$ are thus solely due to surface charges building up.\newline
\indent The probe itself (diameter of \SI{2}{\mm}) is mounted on top of a spherical vacuum chamber and remains in a fixed position at all times, while the copper sample plate is attached to a manipulator allowing for three-dimensional movement of the sample underneath the tip, sketched in Fig.~\ref{fig:Fig2_expSetup}(a). The pressure in the chamber is below \SI{e-8}{\milli\bar}, established by a turbomolecular pump. An area of up to $58 \times 84$ \si{\mm^2} can be scanned and the typical spacing of the tip and the sample is \SIrange[range-phrase = --]{0.2}{2.0}{\milli \metre}. The setup allows for a simultaneous application of two voltages, $\pm\mathrm{V_{applied}}$, up to $\pm$\SI{10}{\kilo\volt} to the positive and negative electrodes while keeping the plate grounded. An integrated heater below the sample plate can heat the substrate up to \SI{250}{\celsius}. This set temperature can be reached within about \SI{30}{\min}.
\newline
\indent To thoroughly characterize the effect of UV light on trapped charges, we use a home-built UV source with tunable wavelength and intensity output. It is based on the broadband metal halide UV lamp \textit{Osram SUPRATREC HTT 150-211} that emits a wide UV spectrum from \SI{250}{\nano\metre} to well into the visual. The lamp has an output power of \SI{6}{\watt} in the UVB (\SIrange[range-phrase = --]{280}{315}{\nano\metre}) and \SI{22}{\watt} in the UVA range of the optical spectrum (\SIrange[range-phrase = --]{315}{400}{\nano\metre}). To ensure stable output, we keep the lamp turned on continuously and use a shutter to control the irradiation of the sample. In the housing of the source, a set of neutral density and band-pass filters can be installed for intensity and wavelength tuning. More details on the home-built UV source and its intensity characterization are provided in Appendix~\ref{sec:UVchar}.\newline 
\indent In our experiment, we employ a  microstructured electrostatic trap,\cite{Eng.11} providing both a box-like potential and high electric fields to trap a large ensemble of cold polar polyatomic molecules and to suppress collisional loss from dipolar relaxation.\cite{Kol.22} Typically, we apply voltages up to  $\pm$\SI{1.6}{\kilo\volt} during operation to the positive/negative electrode of the microstructure and the minimum gap between the electrodes is \SI{130}{\micro\meter}. To achieve this, the structures are coated with a dielectric polymer that is supposed to prevent surface flashover.\cite{Yang.22} \newline 
\indent For less material consumption and rapid investigation, we conduct all experiments with smaller and simplified versions of the microstructures used in the electrostatic molecule trap.\cite{Eng.11} Fig.~\ref{fig:Fig2_expSetup}(b) illustrates the sketch of the ``test'' structure. Apart from the overall dimensions, most other key features are shared between the smaller test structures and their larger counterparts. In both, chromium electrodes of \SI{100}{\nano \meter} thickness are imprinted onto AF32 glass (SCHOTT A.G.) substrate by positive photolithography. Two adjacent electrode strips are at least \SI{175}{\micro\meter} apart. The width of the positive and negative electrodes are \SI{150}{\micro\meter} and  \SI{300}{\micro\meter}, respectively.\newline 
\indent In this study, we use two test structures with identical material and design as described above. However, only one of the two samples is coated with an additional approximately \SI{1} {\micro\metre} thick layer of a dielectric polymer called Cyclotene$^{\text{TM}}$ (by Dow Chemical, XUS35078 type 1) which covers the microstructured region, as in our molecule trap mentioned previously. This coating will have an effect on both the charge buildup and charge removal processes, so a comparison of the samples is likely to unveil information about qualitatively different behaviors between them. For simplicity, we call the samples with and without a Cyclotene$^{\text{TM}}$ layer ``coated sample'' and ``uncoated sample'', respectively.

\section{\label{sec:MeasAna}MEASUREMENT PROCEDURE AND ANALYSIS} 
\noindent The scanning Kelvin probe enables a detailed study of the effects of UV illumination across the whole surface of the charged sample. To facilitate this, we have developed measurement and data-analysis procedures to create reproducible data sets with comparable initial conditions. Moreover, we give an example of a surface charge measurement on the samples after voltage application.

\subsection{Data collection method}\label{subsec:Procedure}
\begin{figure}[t]
\centering
\includegraphics[width=1.0\hsize]{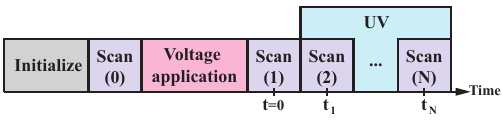}
\caption{\label{fig:Fig4_MeasScheme} Charge-removal-by-UV-illumination measurement procedure. The procedure starts with the sample in the ``initialized'' state, described in the main text. It is scanned once in this state (``Scan (0)'') before high voltage is applied. Immediately after the voltage is switched off, the first scan of the series (``Scan (1)'') is taken. In the following scans, the UV irradiation is kept on continuously and the sample is scanned repeatedly (``Scan (2)'' to ``Scan (N)'') to record the charge removal dynamics. The typical scanning time for an area of $13 \times$ \SI{5}{\milli\metre^2} of a measurement is around \SI{9}{\minute}.} 
\end{figure}

\noindent For each data set, we apply the measurement scheme shown in Fig.~\ref{fig:Fig4_MeasScheme}. A systematic study of charge buildup and charge removal requires a reproducible initial state of the sample. For a newly installed sample, we achieve this state by holding it at \SI{250}{\celsius} for several hours to eliminate adsorbates. After a previous charge-removal test, we return a sample to the initialized state either through successful charge removal by UV illumination or by additionally keeping it at \SI{250}{\celsius} for a couple of hours. We verify this charge-free state before each measurement (called ``Scan (0)'' in Fig.~\ref{fig:Fig4_MeasScheme}). The typical scanning time for an area of $13\times$\SI{5}{\milli\metre^2} (84 points with a step size of \SI{1}{\mm}) is around \SI{9}{\minute}. Subsequently, we apply a voltage of $\pm$$\mathrm{V_{applied}}$ to the positive/negative electrode of the sample. Directly after the voltage is applied and turned off, we take a scan to record the ``charged state'' of the sample before we start UV illumination (``Scan (1)'' in Fig.~\ref{fig:Fig4_MeasScheme}). The sample is then exposed to ultraviolet light continuously while we scan its surface repeatedly over up to \SI{4}{\hour} (``Scan (2)'' to ``Scan (N)'' in Fig.~\ref{fig:Fig4_MeasScheme}). The starting time $(\mathrm{t}=0)$ for each measurement is defined as the time ``Scan (1)'' is halfway completed. Each scan is then labeled by the time of its midpoint, as shown in Fig.~\ref{fig:Fig4_MeasScheme} .\newline

\subsection{Example surface charge buildup upon voltage application}\label{subsec:Buildup}
\begin{figure}[t]
\centering
\includegraphics[width=1\hsize]{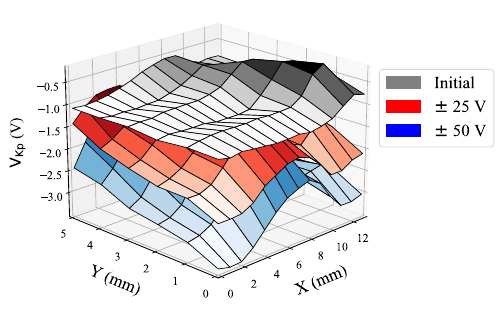}
\caption{\label{fig:Fig3_ChargeBuildUp} An example of surface charge buildup upon voltage application over an area of the sample. Before each measurement, the sample is prepared in an initialized state (refer to Sec.~\ref{subsec:Procedure}; black surface). After voltage application, the $\mathrm{V_{Kp}}$ landscape is altered. With increasing applied voltage, the $\mathrm{V_{Kp}}$ values at different positions all decrease (blue and red surfaces), corresponding to the buildup of more positive surface charges.}
\end{figure}

\noindent Here, we give an example of a typical charge buildup behavior for visualization purposes. We initialize a sample as described in Sec.~\ref{subsec:Procedure}, and then scan it (black surface in Fig.~\ref{fig:Fig3_ChargeBuildUp}). Subsequently, in this specific case, we apply a voltage of $\pm$\SI{25}{\V} to the positive/negative electrode for \SI{1}{\hour} and scan the sample for a second time after turning off the voltage. Doing so drastically changes the shape of the $\mathrm{V_{Kp}}$-surface, as illustrated by the red surface in Fig.~\ref{fig:Fig3_ChargeBuildUp}. For increasing applied voltages, $\mathrm{V_{Kp}}$ becomes more negative and the measured voltages vary more strongly across the surface as shown by the blue surface in Fig.~\ref{fig:Fig3_ChargeBuildUp} $(\pm$\SI{50}{\V}$)$. As a negative $\mathrm{V_{Kp}}$ corresponds to a positive potential difference, positive surface charges must be present on the sample after the voltage application. The data presented in Fig.~\ref{fig:Fig3_ChargeBuildUp} show example charging behavior as a function of applied voltage. 

\subsection{Data analysis and interpretation}\label{subsec:Analysis}

\begin{figure*}
\centering
\includegraphics[width=0.9\hsize]{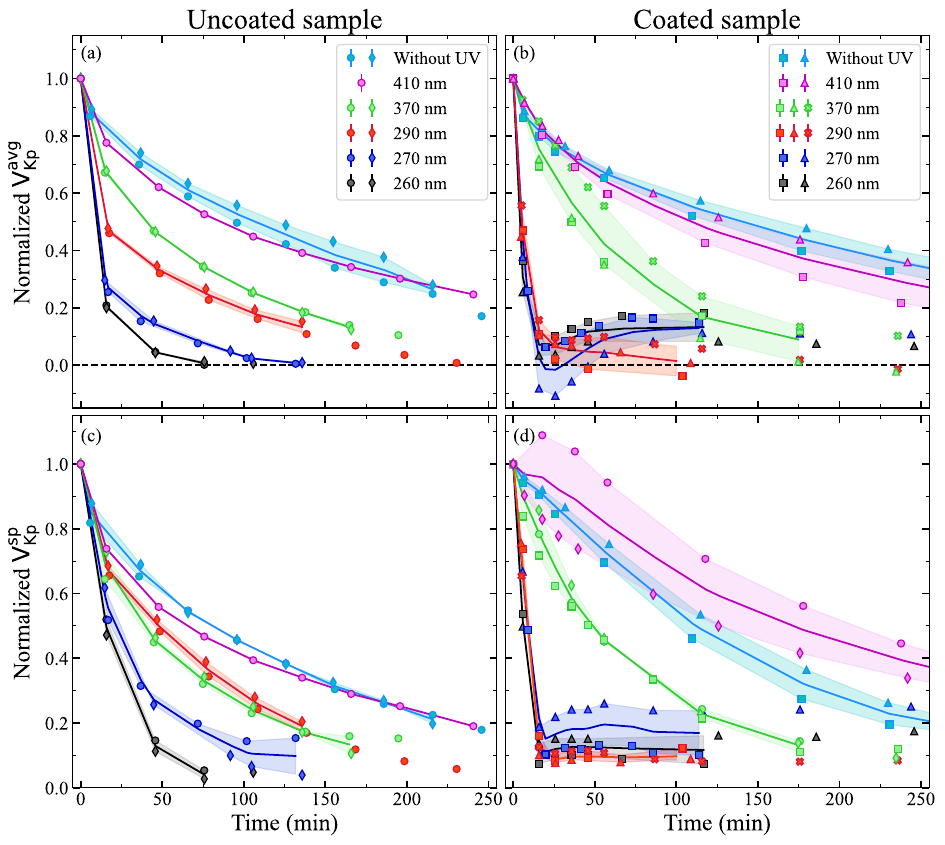}
\caption{\label{fig:Fig5_WavelengthDep} Wavelength dependence of UV-induced charge removal for the uncoated and coated samples. The normalized $\mathrm{V_{Kp}^{avg}}$ and $\mathrm{V_{Kp}^{sp}}$, as defined in Sec.~\ref{subsec:Analysis}, are plotted against UV-illumination time for both the uncoated sample as well as the Cyclotene$^{\text{TM}}$ -coated sample. (a) and (b) show the normalized $\mathrm{V_{Kp}^{avg}}$ for a set of 5 wavelengths for the uncoated and coated samples, respectively. The dashed line indicates a return to the normalized $\mathrm{V_{Kp}^{avg}}$ of the initialized sample. (c) and (d) show the normalized $\mathrm{V_{Kp}^{sp}}$ for the same data. Note that the size of the error bars is small compared with the markers.}
\end{figure*}

\noindent The main concern with surface charges in our experiment is the effects of both the offset and patch potentials caused by them on the homogeneous fields in the center of our electrostatic trap. \cite{Pre.16}  To capture these key aspects, we define two related scalar figures of merit (\textit{average} and \textit{spread}), which are calculated for each individual scan. A single scan $\mathrm{n}$ of the surface consists of $\mathrm{J}$ individual measurements of the Kelvin probe voltage at positions $\mathrm{j}$, $\mathrm{V_{Kp,j}^n}$. The measurement of the initialized sample (``Scan (0)'' in Fig.~\ref{fig:Fig4_MeasScheme}) is labeled $\mathrm{V_{Kp,j}^0}$ and we subtract it pointwise from the Kelvin probe voltages of any scan, $\mathrm{V_{Kp,j}^n}$, yielding the Kelvin probe voltages relative to the initialized state as $\mathrm{V_{Kp,j}^{n,sub}}=\mathrm{V_{Kp, j}^n}-\mathrm{V_{Kp,j}^0}$. After the subtraction, we define the \textit{average} as
\begin{equation}
\mathrm{V_{Kp}^{n, avg}} = \frac{\sum_\text{j}\mathrm{V_{Kp,j}^{n,sub}}}{\text{J}},
\end{equation}
the sum of all $\mathrm{V_{Kp,j}^{n,sub}}$ divided by the number of scanning points $\mathrm{J}$ interpreting as a measure for the overall charge on the surface. We usually find almost all points measured to have the same sign for the voltage. The \textit{spread} of each scan is a measure of this inhomogeneity and is defined as 
\begin{equation}
\mathrm{V_{Kp}^{n,sp} =  \sqrt{\frac{\sum_\text{j} \left(\mathrm{V_{Kp,j}^{n,sub}} - \mathrm{V_{Kp}^{n, avg}} \right)^2}{\text{J}}}},
\end{equation}
the square root of the mean of the quadratic distances of the individual $\mathrm{V_{Kp,j}^{n,sub}}$ from their average $\mathrm{V_{Kp}^{n, avg}}$. For better readability, from now on, we drop the scan index label $\mathrm{n}$ for all quantities defined above. \newline
\indent For an individual $\mathrm{V_{Kp,j}}$, we assign the statistical uncertainty as an estimated standard deviation of $\mathrm{V_{Kp}}$ obtained from calibration measurements, see Appendix~\ref{sec:Uncertainty Determination}. We then obtain the statistical uncertainty of $\mathrm{V_{Kp}^{avg}}$  and $\mathrm{V_{Kp}^{sp}}$ by error propagation.\newline
\indent To compare between measurements, we normalize $\mathrm{V_{Kp}^{avg}}$ and $\mathrm{V_{Kp}^{sp}}$ to ``Scan (1)'' in Fig.~\ref{fig:Fig4_MeasScheme}. This way of processing the scans facilitates the interpretation and comparison of different measurement series: The normalized $\mathrm{V_{Kp}^{avg}}$ will start at 1 and should approach 0 for successful charge removal. If it becomes negative, charges of opposite polarity have been introduced. If it exceeds 1, additional positive charges have been introduced by the UV illumination. Generally, the $\mathrm{V_{Kp}^{sp}}$ can be interpreted in a similar manner. However, it will most likely never reach 0 because of the nature of its definition as the sum of the squared differences from the average and the fact that small variations of the individual $\mathrm{V_{Kp}}$ are present. \newline
\indent In any plots containing such data, we add a trend line as a guide to the eye (see, for example, Fig.~\ref{fig:Fig5_WavelengthDep}). We obtain it by linearly interpolating the data of each measurement series and plotting the mean of those interpolations for the corresponding set of measurement series. The shaded area between the minimum and maximum values among all interpolations gives an idea of the variation between individual measurement series. 

\section{\label{sec:result}RESULTS}
\noindent Following our aim to test the feasibility of UV illumination to reduce offset and patch potentials accumulating on microstructured trap surfaces, we present results for the charge-removal characteristics on both the uncoated and the coated samples according to the procedure outlined in the previous section. Here, we investigate both the wavelength and the intensity dependence of these. We further pay special attention to differences in the behaviors between the uncoated and coated samples to understand the influence of the dielectric coating on our trap structures.

\subsection{Wavelength dependence}
\noindent We recorded charge-removal dynamics at five different wavelengths ranging from \SI{260}{\nano\m} in the UVC to \SI{410}{\nano\m} in the visible for both samples, as shown in Fig.~\ref{fig:Fig5_WavelengthDep}. In addition, we show two data sets without UV illumination (called reference curves) for each sample. For all wavelengths, we keep the intensity of the UV irradiation approximately constant at about \SI{2}{\milli\watt\per\cm^2} (see Appendix~\ref{sec:UVchar}). \newline
\indent For the uncoated sample, we apply $\pm\mathrm{V_{applied}}$ from $\pm$\SI{1700}{\volt} to $\pm$\SI{1850}{\volt} for \SIrange{60}{80}{\minute}, whereas for the coated sample we keep $\pm\mathrm{V_{applied}}$ at $\pm$\SI{1200}{\volt} for \SIrange{2}{6}{\hour}. We vary $\pm\mathrm{V_{applied}}$ and their respective duration in these tests to keep the $\mathrm{V_{Kp}^{avg}}$ after voltage application on the samples constant for each sample individually. This yielded $\mathrm{V_{Kp}^{avg}}$ after voltage application of $-5.3\pm 0.5$ \si{\volt} for the uncoated sample and of $-1.7\pm 0.3$ \si{\volt} for the coated sample. However, there is an indication that the charge-decay dynamics without any UV illumination at short time may not depend on the magnitude of the $\mathrm{V_{Kp}^{avg}}$ as explained in detail in Appendix~\ref{sec:Compare}.\newline
\indent Fig.~\ref{fig:Fig5_WavelengthDep}(a) and (c) show the normalized $\mathrm{V_{Kp}^{avg}}$ and $\mathrm{V_{Kp}^{sp}}$ for the uncoated sample, respectively. A strong wavelength dependence becomes apparent immediately: For the two shortest wavelengths, more than $80\%$ of the charge is removed within the first hour of UV illumination. For the two intermediate wavelengths, charge removal accelerates significantly as well, whereas for \SI{410}{\nano\m} only a small difference to the reference data is apparent. Moreover, the curves for the same wavelengths overlap almost entirely. \newline
\indent Here, possible removal mechanisms of positive surface charges are that positive ions are desorbed from the surface upon irradiation with UV light or that electrons liberated from nearby metal electrodes due to UV illumination neutralize the residual charges on the sample.\cite{Hew.07,Ugo.08} It is noteworthy that the charges disappear slowly from the sample even without UV illumination. This may point to a finite mobility of charge carriers on the insulator part of the sample. Indeed, the slow decay of surface charges on insulators is a known phenomenon for which different mechanisms are proposed. \cite{HAE.75,Kin.08,Gre.13} We cannot distinguish them in our setup due to experimental limitations.\newline
\indent  Fig.~\ref{fig:Fig5_WavelengthDep}(b) and (d) show  the normalized $\mathrm{V_{Kp}^{avg}}$ and $\mathrm{V_{Kp}^{sp}}$  for the coated sample. While the general trends are similar, the data has several features that are very distinct from those of the uncoated sample. The initial charge removal is more rapid for shorter wavelengths than for the uncoated case. Additionally, the normalized $\mathrm{V_{Kp}^{avg}}$ for both \SI{260}{\nano\m} and \SI{270}{\nano\m} exhibits a minimum after about \SI{30}{\minute} of UV application beyond which the normalized $\mathrm{V_{Kp}^{avg}}$ rises again. While one data set for \SI{290}{\nano\m} has a local minimum around \SI{30}{\minute} as well, it remains unknown whether at longer wavelengths similar behavior appears for even longer application times. Similarly, minima can be seen for the normalized $\mathrm{V_{Kp}^{sp}}$. For one measurement series with \SI{270}{\nano\m}, the normalized $\mathrm{V_{Kp}^{avg}}$ becomes even negative for a couple of data points. Moreover, \SI{290}{\nano\m} remove charges almost as rapidly as the two shorter wavelengths. \SI{370}{\nano\m} suffice to enhance charge removal while at \SI{410}{\nano\m}, the behavior is not significantly different from the two reference curves.\newline
\indent There is a noticeable difference between the uncoated and the coated sample regarding the scaling of the effect with the wavelength. While the removal dynamics accelerate gradually towards shorter wavelengths for the former, there is a pronounced step in the latter. In between \SI{370}{\nano\metre} and \SI{290}{\nano\meter}, the charge removal of the coated sample appears to be enhanced significantly. We hypothesize this may be connected to the absorption spectrum of Cyclotene$^{\text{TM}}$ having a steep increase towards shorter wavelengths at this point.\cite{Kao.02} The fact that both the curves for \SI{370}{\nano\meter} as well as for \SI{410}{\nano\meter} overlap for both samples (see Appendix~\ref{sec:Compare}) supports the presumption that the coating will alter the charge-removal process only after it starts absorbing the incident UV irradiation.\newline
\indent Moreover, the variations between measurement series taken for the same wavelength are larger compared to the uncoated sample. Finally, the reference curves for the normalized $\mathrm{V_{Kp}^{avg}}$ of the coated sample drops more slowly at the longer time ($\mathrm{t}>$ \SI{50}{\min}) compared to the uncoated sample in Appendix~\ref{sec:Compare}. While the nature of these effects is uncertain, their existence points to different mechanisms partaking in charge decay on the coated sample.

\begin{figure*}
\centering
\includegraphics[width=0.9\hsize]{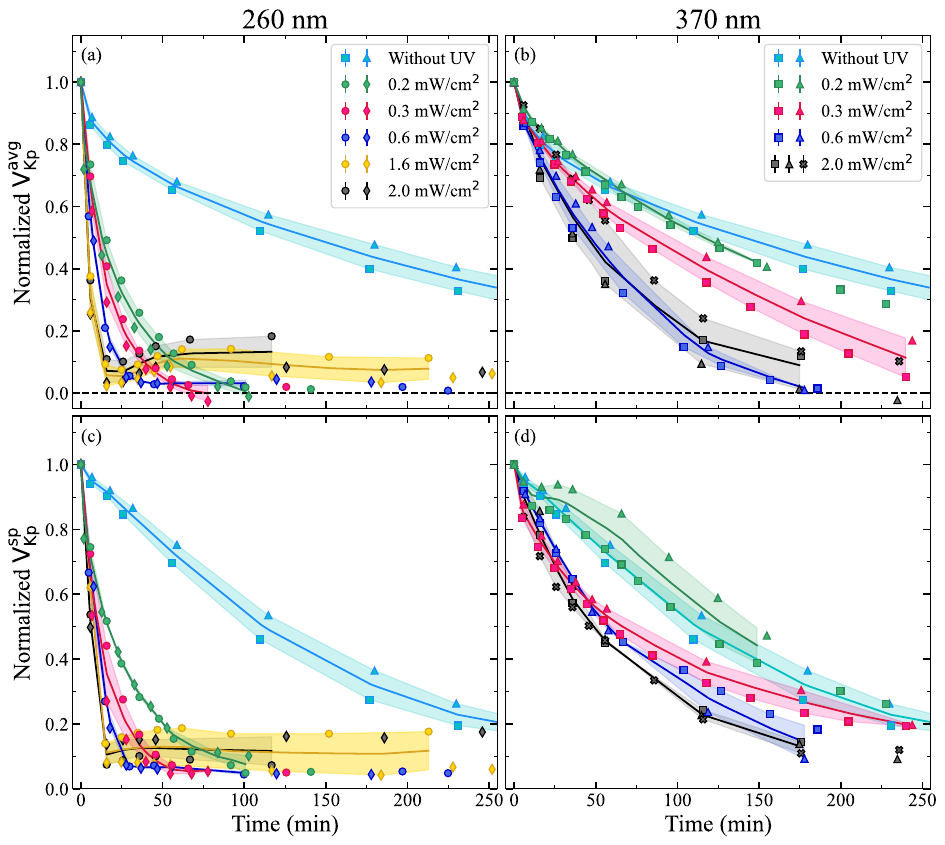}
\caption{\label{fig:Fig6_IntensityDep} Intensity dependence of UV-induced charge removal for the Cyclotene$^{\text{TM}}$ -coated sample for a short and a long UV wavelength. The normalized $\mathrm{V_{Kp}^{avg}}$ and $\mathrm{V_{Kp}^{sp}}$, as defined in Sec.~\ref{subsec:Analysis}, are plotted against UV-illumination time. (a) and (b) show the normalized $\mathrm{V_{Kp}^{avg}}$ depending on the intensities of the UV light for wavelengths at \SI{260}{\nano\m} and \SI{370}{\nano\m}, respectively. The dashed line indicates a return to the normalized $\mathrm{V_{Kp}^{avg}}$ of the initialized sample. (c) and (d) show the normalized $\mathrm{V_{Kp}^{sp}}$ for the same data. Note that the size of the error bars is small compared with the markers.}
\end{figure*}

\subsection{Intensity dependence}
\noindent Fig.~\ref{fig:Fig6_IntensityDep} shows the charge-removal dynamics of the coated sample with intensities ranging from \SI{0.2}{\milli\watt\per\centi\m^2} to \SI{2}{\milli\watt\per\centi\m^2} for \SI{260}{\nano\m} and \SI{370}{\nano\m}. In both cases, the charge-removal speed is higher with higher intensities. Fig.~\ref{fig:Fig6_IntensityDep} (a) and (c) show the normalized $\mathrm{V_{Kp}^{avg}}$ and $\mathrm{V_{Kp}^{sp}}$ at \SI{260}{\nano\m} and these data suggest that the minimum of the values can be reached within \SI{2}{\hour}. Only for the two highest intensities of \SI{1.6}{\milli\watt\per\centi\m^2} and \SI{2.0}{\milli\watt\per\centi\m^2}, there is a pronounced minimum at around \SI{20}{\min}. This feature is absent for all other intensities tested at \SI{260}{\nano\m}. In contrast, for \SI{370}{\nano\m} with lower intensities, the normalized $\mathrm{V_{Kp}^{avg}}$ and $\mathrm{V_{Kp}^{sp}}$  can not reach the initialized potential state within \SI{4}{\hour}, as depicted in Fig.~\ref{fig:Fig6_IntensityDep}(b) and (d).

\section{\label{sec:Conclusion}CONCLUSION}
\noindent Using a scanning Kelvin probe, we confirm electric-potential alterations across the whole surface, consisting of an insulating substrate with microfabricated electrodes on top after voltage application. We then present a systematic investigation of how UV irradiation of the surface helps to reverse these alterations, with special attention being paid to the role of the UV light wavelength and intensity. \newline
\indent We measure a lasting change of the electric potential consistent with positive charges on the sample surface after the application of voltage to the microstructure. Both absolute offset and patch electric potentials on the microstructure increase with the magnitude of the voltage applied. \newline
\indent We have investigated how UV exposure can enhance the removal of offset and patch potentials. Without UV illumination, the surface charge decays down to 20\% of the initial charge after the application of voltage in approximately four hours. For wavelengths from \SI{260}{\nano\meter} to \SI{410}{\nano\meter}, we find all but the longest to increase the speed with which the charges are removed from the sample surface. Moreover, we observe a strict wavelength ordering: The shorter the wavelength, the quicker the charges are removed. For \SI{260}{\nano\meter}, the time it takes for the average charge to drop below 20\% of the initial charge is reduced to about thirty minutes. \newline
\indent We performed the same measurement on a sample coated with a polymer of the Cyclotene$^{\text{TM}}$ family. For the longest UV wavelengths tested (\SI{410}{\nano\meter}), we did not find a difference to the uncoated sample. For wavelengths below this threshold, the charge-removal process is accelerated even further compared to the uncoated sample for the same wavelengths and intensities. We suspect the increase in removal speed is related to the polymer's absorption characteristics.. \newline
\indent Finally, we also investigated the dependence of the charge-removal characteristics on the UV intensity on the coated sample for UV wavelengths of \SI{260}{\nano\meter} and \SI{370}{\nano\meter}. For both, we find higher intensities to enhance the charge removal to a plateau. This plateau is reached at lower intensities for \SI{370}{\nano\meter} than it is for \SI{260}{\nano\meter}. \newline
\indent Especially the longer but still effective wavelengths may be an \textit{in-situ} and simple way to remove surface charges from an insulator due to the availability of high-power LEDs at these wavelengths. This technique may help improve the performance of any devices relying on a geometry similar to our trap.

\begin{acknowledgments}
\noindent This work was supported by Deutsche Forschungsgemeinschaft under Germany's Excellence Strategy via Munich Center for Quantum Science and Technology EXC-2111-390814868. We thank M. Belkin, R. Meyer, L. Mora and G. Riedl for manufacturing the samples investigated in this study. M. Zeppenfeld is acknowledged for fruitful discussions. We thank T. Urban, T. Wiesmeier, F. Furchtsam and J. Siegl for their technical support. We are indebted to Y. Briard for taking measurements and analyzing data at the initial state of the project. J. Phrompao acknowledges funding through stipends from the International Max Planck Research School for Quantum Science and Technology (IMPRS-QST) and from the Development and Promotion of Science and Technology Talents (DPST) jointly administered by the Ministry of Science and Technology, the Ministry of Education, and the Institute for the Promotion of Teaching Science and Technology (IPST) of Thailand.
\end{acknowledgments}

\section*{AUTHOR DECLARATIONS}

\subsection*{Conflict of Interest}
\noindent The authors have no conflicts to disclose. 

\subsection*{Author Contributions}
\noindent M. T. Ziemba and J. Phrompao contributed equally to this work.\newline
\\
\textbf{M. T. Ziemba}: Conceptualization (supporting); Formal analysis (equal); Methodology (lead); Investigation (equal); Resources (equal); Software (lead); Visualization (supporting); Writing - original draft (lead); Writing - review \& editing (equal). \textbf{J. Phrompao}: Conceptualization (lead); Formal analysis (equal); Investigation (equal); Resources (equal); Software (supporting); Visualization (lead); Writing - original draft (supporting); Writing - review \& editing (equal). \textbf{F. Jung}: Formal analysis (equal); Investigation (supporting); Methodology (supporting); Resources (equal); Writing - review \& editing (equal). \textbf{I. M. Rabey}: Conceptualization (equal); Resources (equal); Supervision (supporting); Writing - review \& editing (equal). \textbf{G. Rempe}: Conceptualization (equal); Funding acquisition (lead); Project administration (lead); Resources (equal); Supervision (lead); Writing - review \& editing (equal).

\section*{DATA AVAILABILITY}
\noindent The data that support the findings of this study are available from the corresponding author upon reasonable request.

\appendix

\section{HOME-BUILT UV SOURCE DESIGN AND INTENSITY CHARACTERIZATION}\label{sec:UVchar}

\begin{figure}[b!]
\includegraphics[width=1\hsize]{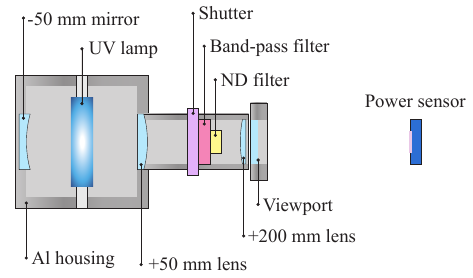}
\caption{\label{fig:UVCharSetup}Home-built UV source and UV-intensity characterization setup. The UV lamp is mounted in an enclosed aluminum housing to prevent UV leakage to the surrounding. A \SI{50}{\mm}-focal-length concave mirror and a \SI{50}{\mm}-focal-length convex lens are used to collect the light from the UV lamp to the output tube. A shutter is installed to switch on and off the UV light. A band-pass filter and/or an ND filter is attached behind the shutter to select a wavelength and intensity of the light, respectively. Another \SI{200}{\mm}-focal-length convex lens is employed to focus the UV light on the sample in the middle of the vacuum chamber. The UV viewport is placed at the output tube and a power sensor is located at the usual distance of the sample around \SI{17}{\mm} away from the UV viewport.}
\end{figure}

\noindent The broadband metal halide UV lamp is mounted in the middle of an aluminum housing as illustrated in Fig.~\ref{fig:UVCharSetup}. For safety reasons and to ensure the stability of the output power, water cooling is employed to keep the housing temperature at less than \SI{50}{\celsius} as the bulb temperature reaches \SI{650}{\celsius} to \SI{850}{\celsius} during its operation. To collect the light from the lamp, a \SI{50}{\mm}-focal-length concave mirror and a \SI{50}{\mm}-focal-length convex lens are installed behind and in front of the bulb. The UV lamp needs to be switched on for at least 2 minutes to reach its full output power. A shutter allows continuous operation of the UV source without switching the lamp on and off. A band-pass filter with optional neutral-density filters is attached to the shutter, allowing the choice of wavelength and output power. The FWHM of the band-pass filters is $10\pm 2$ \si{\nm}. Finally, a \SI{200}{\mm}-focal-length lens is used to focus the UV light onto the sample. The UV source is attached to a UV viewport via which the UV light illuminates the sample at a \ang{45} angle to the surface of the sample.

\begin{table}[b!]
\caption{\label{tab:table1}UV light intensities for surface-charge removal by UV illumination \textit{wavelength dependence} for five different wavelengths as shown in Fig.~\ref{fig:Fig5_WavelengthDep}. Note that the uncertainties of the intensities are the standard deviations.}

\begin{ruledtabular}
\begin{tabular}{llc}
Wavelength (\si{\nano \meter})& OD filter&Intensity (\si{\milli \watt / \cm^2})\\
\hline
 410&  -&$1.86\pm 0.05$\\
370&  -&$1.87\pm 0.06$\\
290&  0.4&$1.77\pm 0.06$\\
 270&  0.5&$1.86\pm 0.07$\\
 260&  0.5&$1.95\pm 0.05$\\
\end{tabular}
\end{ruledtabular}
\end{table}

\begin{table}[b!]
\caption{\label{tab:table2}UV light intensities for surface-charge removal by UV illumination \textit{intensity dependence} as shown in Fig.~\ref{fig:Fig6_IntensityDep}. Note that the uncertainties of the intensities are the standard deviations.}
\begin{ruledtabular}
\begin{tabular}{llc}
Wavelength (\si{\nano \meter})& OD filter&Intensity (\si{\milli \watt / \cm^2})\\
\hline
 260&  1.6&$0.161\pm 0.004$\\
260&  1.3&$0.332\pm 0.009$\\
260&  1.0&$0.632\pm 0.019$\\
 260& 0.6&$1.60\pm 0.05$\\
 260&  0.5&$1.95\pm 0.05$\\
 370&  0.8&$0.151\pm 0.004$\\
 370& 0.4&$0.332\pm 0.009$\\
 370& 0.1&$0.620\pm 0.021$\\
 370& -&$1.87\pm 0.06$\\
\end{tabular}
\end{ruledtabular}
\end{table}

\indent To measure the intensity of the UV light illuminating the sample, it is not practical to measure at the sample position in the UHV chamber. Instead, the UV source, including the UV viewport, was tested on an optical table and the intensity at the sample distance was measured. The only difference to the real setup is that the incident light is perpendicular to the aperture of the power meter. A Thorlabs sensor S120VC with an aperture size of \SI{9.5}{\mm} is used for the intensity characterization. For each wavelength and a set of neutral-density filters, we measure the power 1000 times, which takes around 2 minutes. Table~\ref{tab:table1} presents the intensity of the UV light employed in the wavelength dependence measurement as shown in Fig.~\ref{fig:Fig5_WavelengthDep}. The intensities of the different wavelengths of the UV light are comparable, most of them agree within their respective standard deviations. Table~\ref{tab:table2} shows the intensity of the UV light for the intensity dependence measurement for the  \SI{260}{\nm} and  \SI{370}{\nm} wavelengths as illustrated in Fig.~\ref{fig:Fig6_IntensityDep}. Furthermore, long-term power stability of the output wavelengths of \SI{290}{\nm} with a 0.4 OD filter and \SI{370}{\nm} without OD filter are observed for \SI{4}{\hour}. The intensities of \SI{260}{\nano\m} and \SI{370}{\nano\m} are $2.04\pm 0.09$ \si{\milli \watt / \cm^2} and $1.97\pm 0.11$ \si{\milli \watt / \cm^2}, respectively. Note that the uncertainties of the intensities are the standard deviations.

\section{UNCERTAINTY DETERMINATION OF AN INDIVIDUAL $\mathrm{V_{Kp}}$ }\label{sec:Uncertainty Determination}

\begin{figure*}
\centering
\includegraphics[width=1\hsize]{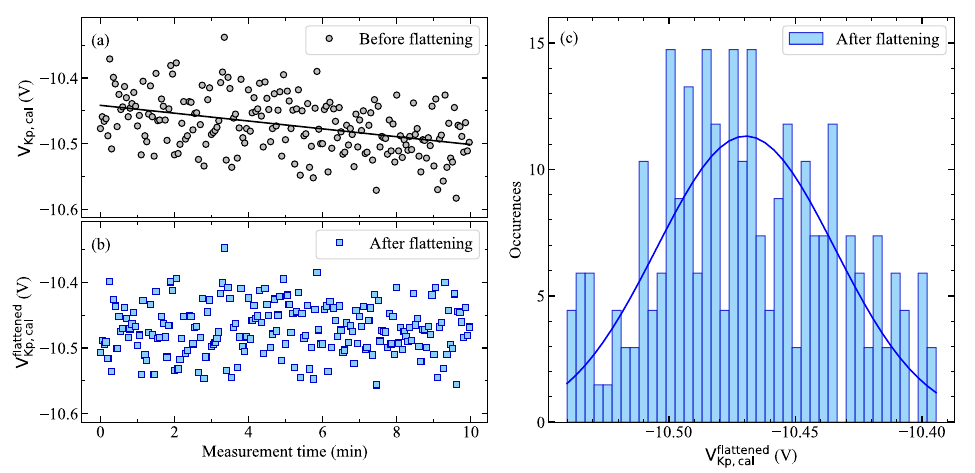}
\caption{\label{fig:Fig7_Flattening} The calibration procedure was demonstrated on an exemplary data set. (a) The gray circles show measured $\mathrm{V_{Kp,cal}}$, with the non-zero slope indicating charging during the measurement. The black line is a linear fitting function that is used for the flattening procedure. (b) The blue squares show the $\mathrm{V_{Kp,cal}^{flattened}}$ after the flattening procedure has been performed. (c) Histogram of $\mathrm{V_{Kp,cal}^{flattened}}$, with a normal distribution function as a guide to the eye (illustrated in the blue curve).} 
\end{figure*}

\noindent The uncertainty of an individual $\mathrm{V_{Kp,j}}$ is assigned as an estimated standard deviation of the $\mathrm{V_{Kp}}$ obtained from calibration measurements. These measurements are reproducing typical $\mathrm{V_{Kp}}$ (ranging from \SI{-11.4}{\volt} to \SI{0.7}{\volt}) as taken from charge-removal-by-UV-illumination measurements described in Sec.~\ref{sec:MeasAna}. Differently from the actual measurement scheme introduced in Sec.~\ref{sec:MeasAna}, the calibration Kelvin probe Voltages $\mathrm{V_{Kp,cal}}$ are here produced by applying a constant voltage to one electrode of the sample and keeping the other electrode grounded while measuring with the Kelvin probe. This method is chosen because it allows us to determine the uncertainty of $\mathrm{V_{Kp}}$, $\mathrm{\sigma_{V_{Kp}}}$, as a function of $\mathrm{V_{Kp}}$ without needing to charge the sample surface by high-voltage application over several hours. The measured voltage is independent of how it is produced, whether caused by an applied voltage or by the presence of surface charges. In addition, voltages of this size can be applied with the probe \textit{in-situ} without risking the destruction of the Kelvin probe by breakdown between the probe tip and the sample.

\begin{figure}[t!]
\centering
\includegraphics[width=1\hsize]{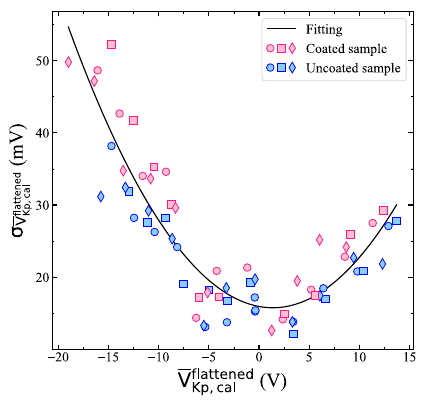}
\caption{\label{fig:Fig8_CalibrationData}Calibration data and curve for the uncertainty determination of an individual $\mathrm{V_{Kp}}$. The blue and pink markers show the $\mathrm{\sigma_{\overline{ V}_{Kp,cal}^{flattened}}}$ as a function of $\mathrm{\overline{V}_{Kp,cal}^{flattened}}$ for uncoated and coated samples, respectively. The black curve is a quadratic function fitted to the data sets. It is then used to assign the uncertainty to an individual $\mathrm{V_{Kp}}$.}
\end{figure}

\indent For each calibration measurement, one position on a sample is measured 200 times while applying a constant voltage to the sample throughout the whole measurement. This application of voltage also leads to the charging of the insulator, altering the $\mathrm{V_{Kp,cal}}$ as a function of time. However, the contribution of the introduced charges to $\mathrm{V_{Kp,cal}}$ remains small (roughly a few hundred millivolts per 10-minute measurement), such that calibration curves at different applied static voltages are clearly separated and the small alteration can be taken into account by a flattening procedure. A typical example of the $\mathrm{V_{Kp,cal}}$ obtained in a single calibration measurement is shown in Fig.~\ref{fig:Fig7_Flattening} (a) where \SI{20}{\volt} were applied to the sample, showing the charging effect as an overall decrease in $\mathrm{V_{Kp,cal}}$ over time. To account for this charging effect in the analysis, we perform a flattening procedure by first extracting the linear trend from a linear function fit $\mathrm{V_{Kp,cal}(t)}=a \mathrm{t}+b$, where $ \mathrm{t}\in [0,\mathrm{t_{max}}]$, and then transforming the data by $\mathrm{V_{Kp,cal}^{flattened}(t)}=\mathrm{V_{Kp,cal}(t)}-a \mathrm{t}^{'}$, where $\mathrm{t^{'}}=\mathrm{t}-\frac{\mathrm{t_{max}}}{2}$.\newline
\indent As it is apparent from Fig.~\ref{fig:Fig7_Flattening} (b), the variance of the flattened data, which is an estimate for the standard deviation in the case if there were no charging during voltage application, is in a very good approximation independent of the measurement time. It thus also shows no dependence on the different $\mathrm{V_{Kp,cal}^{flattened}}$ induced by the charging effect. This also demonstrates that the flattening procedure was justified. Fig.~\ref{fig:Fig7_Flattening} (c) shows the histogram after the flattening and outlier removal, from which the mean value and standard deviation estimate can be extracted to obtain a $\mathrm{\overline{V}_{Kp,cal}^{flattened}}$ and its corresponding statistical uncertainty $\mathrm{\sigma_{\overline{ V}_{Kp,cal}^{flattened}}}$.\newline
\indent To confirm that the uncertainty is independent of the positions and types of samples, measurements are performed at three different points each on both coated and uncoated samples. In addition, we mimic differently charged states by applying different static voltages in random order. Fig.~\ref{fig:Fig8_CalibrationData} shows the standard deviation $\mathrm{\sigma_{\overline{ V}_{Kp,cal}^{flattened}}}$ as a function of $\mathrm{\overline{V}_{Kp,cal}^{flattened}}$ for three different points on each of the two samples. It is apparent that the standard deviation of the two types of samples has the same trend and is independent of the measured positions for each individual sample. The black curve is a quadratic function used to fit with all data points as illustrated in Fig.~\ref{fig:Fig8_CalibrationData}. Finally, the fitting function is used as the calibration function for the uncertainty estimate of an individual $\mathrm{V_{Kp}}$, and the uncertainty of the $\mathrm{V_{Kp}^{avg}}$ and $\mathrm{V_{Kp}^{sp}}$ are obtained by error propagation. 

\section{COMPARISON OF CHARGE-DECAY DYNAMICS OF THE TWO SAMPLES}\label{sec:Compare}
\begin{figure*}
\centering
\includegraphics[width=1\hsize]{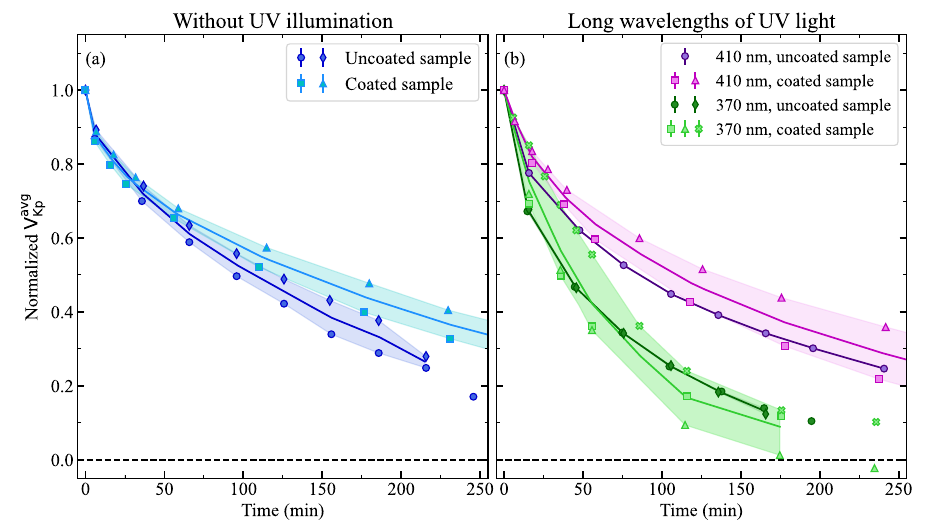}
\caption{\label{fig:CompareDynamicsTwoSamples} Charge-decay curves of uncoated and coated samples with and without long-wavelength-UV illumination. The normalized $\mathrm{V_{Kp}^{avg}}$ and $\mathrm{V_{Kp}^{sp}}$, as defined in Sec.~\ref{subsec:Analysis}, are plotted against time. (a) Charge-decay curves without UV illumination overlap at short times within the blue and turquoise bands. (b) Charge-decay curves with \SI{370}{\nm} UV illumination also overlap within the light and the dark green bands. For the \SI{410}{\nm} UV illumination, the violet line is slightly under the magenta bands. The dashed line indicates a return to the normalized $\mathrm{V_{Kp}^{avg}}$ of the initialized sample. Note that the size of the error bars is small compared with the markers.}
\end{figure*}

\noindent Fig.~\ref{fig:CompareDynamicsTwoSamples}(a) shows the comparison of the charge-decay dynamics between the two samples without UV illumination where the dashed lines indicate a return to the normalized $\mathrm{V_{Kp}^{avg}}$ of the initialized sample. Despite the initial charge of the two samples being different, the charge-decay curves overlap within the blue and the turquoise bands at times shorter than \SI{50}{\minute}. Additionally, the charge-decay curves of the two samples with \SI{370}{\nm} UV illumination are also overlapping within the light and the dark green bands as shown in Fig.~\ref{fig:CompareDynamicsTwoSamples}(b). For the charge-decay curves of the two samples with \SI{410}{\nm} UV illumination, the violet line is slightly below the magenta bands.

\end{document}